\documentclass[a4paper,conference]{IEEEtran}
\IEEEoverridecommandlockouts

\newcommand\hmmax{0}
\newcommand\bmmax{0}

\usepackage{graphicx}
\usepackage{tikz}
\usepackage{pgfplots}
\usepackage{xcolor}
\usepackage{color, colortbl}
\usepackage{amsmath}
\usepackage{bbm}
\usepackage{amsfonts, amssymb}
\usepackage{mathtools}
\usepackage{cite}
\usepackage[nolist]{acronym}

\usepackage{booktabs} 		
\usepackage{diagbox}		
\usepackage{upgreek}
\usepackage{bbm}
\usepackage{Files/bm}
\usepackage{Files/winsnotation}
\usepackage{Files/Symbols_OPL}
\usepackage{Files/LVcolours} 

\usepackage[ruled,vlined,noresetcount]{algorithm2e}
\usepackage{setspace}	 	
\usepackage{comment}
\usepackage{physics}

\usepackage{array}
\newcolumntype{C}[1]{>{\centering\arraybackslash}p{#1}}

\makeatletter
\def\endthebibliography{%
  \def\@noitemerr{\@latex@warning{Empty `thebibliography' environment}}%
  \endlist
}
\makeatother
\usepackage{amsthm}
\theoremstyle{definition}

\newcommand{\px}{p_{\mathrm{X}}}
\newcommand{\py}{p_{\mathrm{Y}}}
\newcommand{\pz}{p_{\mathrm{Z}}}

\newcommand{\PauliZ}{\M{Z}}

\newcommand{\PauliXped}[1]{\M{X}_{{#1}}}
\newcommand{\PauliYped}[1]{\M{Y}_{{#1}}}

\newcommand{\eg}{e_{\mathrm g}}
\newcommand{\eZ}{e_{\mathrm Z}}
\newcommand{\Syndrome}[0]{\V{s}}

\pgfplotsset{compat=1.17}

\begin{document}

\title{Performance Analysis of Quantum Error-Correcting Surface Codes over Asymmetric Channels}


\author{%
  \IEEEauthorblockN{Lorenzo Valentini, Diego Forlivesi, Marco Chiani}
  \IEEEauthorblockA{CNIT/WiLab, DEI, University of Bologna, Italy \\
  	Email: \{lorenzo.valentini13, diego.forlivesi2, marco.chiani\}@unibo.it }
}

\maketitle 

\begin{acronym}
\small
\acro{AWGN}{additive white Gaussian noise}
\acro{BCH}{Bose–Chaudhuri–Hocquenghem}
\acro{CDF}{cumulative distribution function}
\acro{CRC}{cyclic redundancy code}
\acro{LDPC}{low-density parity-check}
\acro{ML}{maximum likelihood}
\acro{MWPM}{minimum weight perfect matching}
\acro{QECC}{quantum error correcting code}
\acro{PDF}{probability density function}
\acro{PMF}{probability mass function}
\acro{MPS}{matrix product state}

\end{acronym}
\setcounter{page}{1}

\begin{abstract}
One of the main challenge for an efficient implementation of quantum information technologies is how to counteract quantum noise. Quantum error correcting codes are therefore of primary interest for the evolution towards quantum computing and quantum Internet. We here analyze the performance of surface codes, one of the most important class for practical implementations, on both symmetric and asymmetric quantum channels. We derive approximate expressions, confirmed by simulations, to evaluate the performance of surface codes and of XZZX codes, and provide a metric to assess the advantage of codes with respect to uncoded systems. Our findings allow to characterize the performance by means of analytical formulas of surface codes, like, for example, the $[[13,1,3]]$, the $[[23,1,3/5]]$, the $[[33,1,3/7]]$, and the  $[[41,1,5]]$ surface codes. 
\end{abstract}


\section{Introduction}

The exploitation of the unique features of quantum mechanics has opened new perspectives on how we can sense, process, and communicate information \cite{Kim:08,Pre:18,OSW:18,NAP:19,WehElkHan:18,CacCalVan:20}. From an engineering point of view there are many challenges to solve, calling for both theoretical and experimental research studies. The aim is to progress towards the already known possible applications of quantum information technologies, as well as those currently still unforeseen, that will arise when  practical implementations becomes available. 
One of the main challenge is how to deal with the noise caused by unwanted interaction of the quantum information with the environment \cite{Sho:95,Laf:96,Got:96,Kni:97,FleShoWin:08,FleShoWin:08a,NieChu:10,RehShi:21, ZorDePGio:23}. Quantum error correcting codes, where a redundant representation of quantum states protects from certain types of errors, are therefore of paramount importance for quantum computation, quantum memories and quantum communication systems \cite{Sho:95,Laf:96,Got:96,Kni:97,FleShoWin:08,FleShoWin:08a,Got:09,Ter:15,MurLiKim:16,Rof:19,Bab:19,ChiConWin:20,NieChu:10,RehShi:21,ZorDePGio:23}. 

In this paper we analyze the performance of surface codes, one of the most important class for practical implementations \cite{BraKit:98,FowMarMar:12, AchRajAle:22, KriSebLac:22, ZhaYouYe:22}. Despite their importance, the performance of these codes has been investigated in the literature only partially, and mainly in terms of accuracy threshold over symmetric channels \cite{BraKit:98,FowMarMar:12,BraSerSuc:14,AtaTucBar:21}. Here we provide a thorough investigation of codes with rates greater than $1/41$, considering symmetric surface codes, asymmetric surface codes, and XZZX codes, on quantum channels with both symmetric and asymmetric errors (depolarizing and polarizing channels). 



The key contributions of the paper can be summarized as follows:
\begin{itemize}
    \item we analyze the performance of surface codes over the depolarizing channel;
    \item we analyze the performance of surface codes over asymmetric channels;
    \item we analytically derive asymptotic approximations of the code performance and thresholds, defining the code-effective threshold of a quantum code;
    \item we provide a comparison between analysis and simulation results obtained with  \acf{MPS} or \acf{MWPM} decoders, in particular for the $[[13,1,3]]$, the $[[23,1,3/5]]$, the $[[33,1,3/7]]$, and the  $[[41,1,5]]$ surface codes. 
\end{itemize}
%



This paper is organized as follows. Section~\ref{sec:preliminary} introduces preliminary concepts and models together with some background material. Section~\ref{sec:analysis} provides the analytical investigation of surface codes. Numerical results are shown in Section~\ref{sec:NumRes}. 

\section{Preliminaries and Background}
\label{sec:preliminary}

\subsection{Quantum Error Correction}
A qubit is an element of the two-dimensional Hilbert space $\mathcal{H}^{2}$, with basis $\ket{0}$ and $\ket{1}$ \cite{NieChu:10}. 
The Pauli operators $\M{I}, \M{X}, \M{Z}$, and $\M{Y}$, are defined by  $\M{I}\ket{a}=\ket{a}$, $\M{X}\ket{a}=\ket{a\oplus 1}$, $\M{Z}\ket{a}=(-1)^a\ket{a}$, and $\M{Y}\ket{a}=i(-1)^a\ket{a\oplus 1}$ for $a \in \lbrace0,1\rbrace$. These operators either commute or anticommute with each other. 
We indicate with $[[n,k,d]]$ a \ac{QECC} that encodes $k$ information qubits  $\ket{\varphi}$ (called logical qubits), into a codeword of $n$ qubits  $\ket{\psi}$ (called data qubits), able to correct all patterns up to $t = \lfloor(d-1)/2 \rfloor$ errors (and some patterns of more errors). The codewords will be assumed equiprobable in the following. 
We use the stabilizer formalism, where a stabilizer code $\mathcal{C}$ is generated by $n-k$ independent and commuting operators $\M{G}_i \in \mathcal{G}_n$, called generators, where $\mathcal{G}_n$ is the Pauli group on $n$ qubits \cite{Got:09,NieChu:10}.   
The code $\mathcal{C}$ is the set of quantum states $\ket{\psi}$ satisfying $\M{G}_i \ket{\psi}=\ket{\psi},\, i=1, 2, \ldots, n-k$. 
Assume a codeword $\ket{\psi} \in \mathcal{C}$ affected by a channel error. 
Measuring the codeword according to the stabilizers $\M{G}_i$ with the aid of ancilla qubits, the error collapses on a discrete set of possibilities represented by combinations of Pauli operators $\M{E}\in \mathcal{G}_n$ \cite{Got:09}. 
Thus, an error can be described by specifying the single Pauli errors on each qubit. 
The measurement procedure over the ancilla qubits results in a quantum error syndrome $\Syndrome(\M{E})=(s_1, s_2, \ldots,s_{n-k})$, with each $s_i =0$ or $1$ depending on $\M{E}$ commuting or anticommuting with $\M{G}_i$, respectively \cite{Got:09}. 
A maximum likelihood decoder will then infer the most probable error $\M{\hat{E}}\in \mathcal{G}_n$ compatible with the measured syndrome.  
The weight of an error is the number of single Pauli errors ($\M{X}$, $\M{Z}$, and $\M{Y}$) occurred on a codeword. For example, the error $\M{E} = \PauliXped{2} \PauliYped{3}$ means that $\M{X}$ is occurred on the second qubit, $\M{Y}$ is occurred on the third qubit, and the weight is two.
A simple channel model is the one characterized by errors occurring independently and with the same statistic on the individual qubits of each codeword. In this model, the error can be $\M{X}$, $\M{Z}$ or $\M{Y}$ with probabilities $\px$, $\pz$, and $\py$, respectively. The probability of a qubit generic error is $\rho = \px + \pz + \py$.
Two important models are the \emph{depolarizing channel} where $\px = \pz = \py = \rho / 3$, and the \emph{phase-flip channel} where $\rho=\pz$, $\px = \py=0$. In the following, we will also adopt the notation $[[n, k,d_\mathrm{X}/d_\mathrm{Z}]]$ for asymmetric codes able to correct all patterns up to $t_\mathrm{X} = \lfloor(d_\mathrm{X}-1)/2\rfloor$ Pauli $\M{X}$ errors and $t_\mathrm{Z} = \lfloor(d_\mathrm{Z}-1)/2\rfloor$ Pauli $\M{Z}$ errors.

\subsection{Theoretical Performance Analysis of Quantum Codes}
The codeword error probability, referred in the following also with the term logical error probability, for a standard $[[n,k]]$ \ac{QECC} which corrects up to $t$ generic errors per codeword, is upper bounded by
\begin{equation}
\label{eq:PeGen}
\rho_\mathrm{L} = 1 - \sum_{j = 0}^{t} \binom{n}{j} \rho^j \, (1-\rho)^{n-j}
\end{equation}
where $\rho=\px+\py+\pz$ is the qubit error probability. This bound assumes a decoding failure whenever the total number of qubits in error exceeds $t$.  
The logical error probability analysis has been recently generalized to $[[n,k]]$ \acp{QECC} which correct up to $\eg$ generic errors and up to $\eZ$ Pauli $\PauliZ$ errors per codeword, over an asymmetric channel with arbitrary  $\px$, $\py$ and $\pz$  \cite{ChiVal:20a}. %
In particular, by weighting each pattern with the corresponding probability of occurrence, the expression \eqref{eq:PeGen} becomes \cite{ChiVal:20a}
\begin{align}
\label{eq:PeGen2}
\rho_\mathrm{L} &= 1 - \sum_{j = 0}^{\eg+\eZ} \binom{n}{j}(1-\rho)^{n-j}\sum_{i = (j-\eg)^+}^{j}\binom{j}{i} \pz^i \left(\px+\py\right)^{j-i} \\
&=1 - \sum_{j = 0}^{\eg+\eZ} \binom{n}{j}(1-\rho)^{n-j}\sum_{i = (j-\eg)^+}^{j}\binom{j}{i} \, \pz^i \, \left(\rho-\pz\right)^{j-i}
\end{align}
where $(x)^+=\max\{x,0\}$. In the case of asymmetric channels with asymmetry described by the bias parameter $A = 2\pz /(\rho - \pz)$, the expression in \eqref{eq:PeGen2} can be simplified to 
\begin{align}
\label{eq:PeAsym}
\rho_\mathrm{L} =&\: 1 - \sum_{j = 0}^{\eg+\eZ} \alpha_j \binom{n}{j}\rho^j  (1-\rho)^{n-j}
\end{align}
where
\begin{align}
\label{eq:PeGenxi_2}
\alpha_j = 
&\begin{dcases}
1 & \text{if}~j\le\eg\\ 
\left(\frac{2}{A+2} \right)^j \sum_{i = j-\eg}^{j}\binom{j}{i} \left( \frac{A}{2} \right) ^i & \text{otherwise} . \,\\
\end{dcases} 
\end{align} 

\subsection{Quantum Topological Codes}

One of the most interesting family of \ac{QECC} is the one regarding \emph{topological} codes. The general design principle behind these codes is that they are built by patching together repeated elements. Using this kind of approach, they can be easily scaled in size in order to raise the distance of the code, and stabilizers commutativity is ensured. As regards the actual implementation, these codes have a great intrinsic advantage. In fact, they require only nearest-neighbor interactions. This is an important property since it seems difficult, for current quantum implementations of encoders and decoders, to perform high-fidelity long-range operations between qubits\cite{Rof:19}.

\begin{figure}[t]
	\centering
	\resizebox{0.50\textwidth}{!}{
	    \input{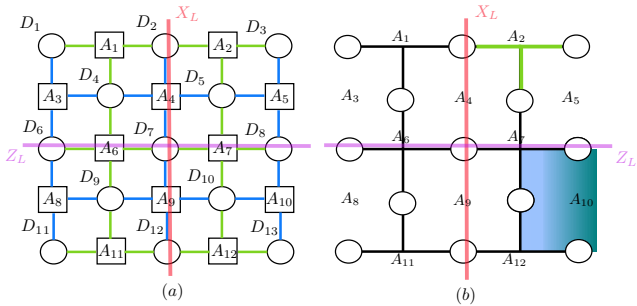}
	}
	\caption{$[[ 13,1,3 ]]$ Surface code. (a) Actual lattice with data qubits (circles) and ancillas (squares). (b) Simplified representation with $\M{X}$ stabilizers corresponding to sites and $\M{Z}$ stabilizers to plaquettes. A smooth edge and a rough edge are depicted in green and blue, respectively. Examples of logical operators are drawn on the lattice. }
	\label{Fig:surface}
\end{figure}

The most important codes within this category are the \emph{surface} codes, in which all the check operators are local and the qubits are arranged on planar sheets. Each stabilizer is associated either with one of the sites, or one of the cells that are called \lq\lq plaquettes\rq\rq. In this structure, logical qubits are defined using spare degrees of freedom in the lattice \cite{HorFowDev:12}. A common way of doing so, consists in defining a lattice having boundaries with two different kinds of edges. In this case, check operators in the interior are four-qubits plaquette or site operators, while the ones at the boundaries are three-qubits operators.  Along a plaquette or \lq\lq rough\rq \rq \; edge, each stabilizer is  a three-qubits operator $\M{Z}^{\otimes 3}$, while, along a site edge or \lq\lq smooth\rq\rq \; edge, each check operator is a three-qubits operator $\M{X}^{\otimes 3}$. Using this procedure, only a single degree of freedom is introduced, hence the entire lattice is able to encode just one logical qubit.
It can be shown that in a code with distance $d$, the lattice has $d^2 + (d-1)^2$ links, which correspond to physical qubits \cite{DenKitLan:02}. For example, two equivalent graphical representations of the resulting lattice for the $[[13,1,3]]$ surface code are shown in Fig.~\ref{Fig:surface}. For this surface code the  stabilizers are
%
%
\begin{equation*}
\arraycolsep=2.5pt
\begin{array}{lllllll} 
\M{G}_1 = \mathbf{X}_1 \mathbf{X}_2 \mathbf{X}_4 &
\M{G}_2 = \mathbf{X}_2 \mathbf{X}_3 \mathbf{X}_5 & \\
\M{G}_3 = \mathbf{Z}_1 \mathbf{Z}_4 \mathbf{Z}_6 &
\M{G}_4 = \mathbf{Z}_2 \mathbf{Z}_4 \mathbf{Z}_5 \mathbf{Z}_7 &
\M{G}_5 = \mathbf{Z}_3 \mathbf{Z}_5 \mathbf{Z}_8 \\
\M{G}_6 = \mathbf{X}_4 \mathbf{X}_6 \mathbf{X}_7 \mathbf{X}_9 &
\M{G}_7 =  \mathbf{X}_5 \mathbf{X}_7 \mathbf{X}_8 \mathbf{X}_{10} & \\
\M{G}_8 = \mathbf{Z}_6 \mathbf{Z}_9 \mathbf{Z}_{11} &
\M{G}_9 = \mathbf{Z}_7 \mathbf{Z}_9 \mathbf{Z}_{10} \mathbf{Z}_{12} &
\M{G}_{10} = \mathbf{Z}_8 \mathbf{Z}_{10} \mathbf{Z}_{13} \\
\M{G}_{11} = \mathbf{X}_9 \mathbf{X}_{11} \mathbf{X}_{12} &
\M{G}_{12} = \mathbf{X}_{10} \mathbf{X}_{12} \mathbf{X}_{13} \,. &
\end{array} 
\end{equation*} 
It is necessary to remind that applying a Pauli operator that commutes with all the stabilizers will preserve the code space. For instance, the tensor product of $\M{Z}$'s and $\M{I}$'s is considered. This operator trivially commutes with all the plaquettes, but it commutes with a site operator only if an even number of $\M{Z}$'s acts on the the links adjacent to the site. A dual reasoning can be done for the $\M{X}$'s and $\M{I}$'s operators. In general, a Pauli operator that commutes with all the stabilizers can be represented as a tensor product of $\M{Z}$'s acting on a closed loop of the lattice, called \emph{cycle}, times a tensor product of $\M{X}$'s acting on a cycle in the dual lattice (where sites and plaquettes have exchanged definitions). 
In this lattice, the links on which these $\M{Z}$'s operators act can form an open path, which begins and ends on rough edges. Reminding that the boundaries of surface codes are not periodic, it follows that these chains of operators are not closed loops and they can be distinguished into two varieties. The first case is the one in which this cycle can be tiled with plaquettes, considering the boundary plaquette too. In this situation the product of $\M{Z}$'s is contained in the stabilizer. Contrarily, a cycle that begins and ends on a rough edge, crossing the lattice, commutes with all the check operators but it is not contained in the stabilizer. As a result, this operator maps the codeword into a different one. A dual reasoning can be done  considering $\M{X}$'s chains on the dual lattice.  
The definition of the logical $\M{X}_L$ and $\M{Z}_L$ operators for the code is straightforward. In particular, the logical $\M{Z}_L$ operator can be chosen as a tensor product of $\M{Z}$'s acting on a chain of qubits running from a rough edge to the one at the opposite side of the lattice. Similarly, the $\M{X}_L$ logical operator will be the tensor product of $\M{X}$'s acting on a chain running from a smooth link to the correspondent one at the other side of the dual lattice. As an example, the 13 surface code, with distance 3, is shown in Fig.~\ref{Fig:surface}.
Recently, a new kind of topological codes has been designed, called XZZX surface codes \cite{AtaTucBar:21}. The XZZX code has the same structure of the surface code, but it differs by a Hadamard rotation on alternate qubits. The basic idea behind these codes is to exploit some symmetries of $\M{X}$ and $\M{Z}$ errors. In particular, a disadvantage of the surface codes is that there are half of the check operators looking for $\M{X}$ errors, and the other half searching for $\M{Z}$ errors. As a consequence, each ancilla has to deal with possible errors of only one type but in four different qubits. Hence, the aim of these new codes is to make each stabilizer equally responsible. In fact, each check operator looks for $\M{X}$ errors on two qubits, and for $\M{Z}$ errors on two different qubits. This fact provides a great advantage in the decoding over asymmetric channels. Indeed, it can be shown that $\M{X}$ and $\M{Z}$ can align only onto one direction, and these two directions are orthogonal to each other. One of the possible decoding techniques for these codes is the \ac{MPS} decoder \cite{BraSerSuc:14}. 

\section{Analysis of Surface Codes}
\label{sec:analysis}
 In addition to the previous properties, surface codes have a great advantage: the value of their accuracy threshold is 0.11\cite{DenKitLan:02}. This is important because, according to the threshold theorem, the logical error rate can be lowered at will if the qubits physical error rate is below this threshold\cite{Rof:19}. 
Another important feature is that a simple decoder exists for such codes: the \ac{MWPM}. This decoder finds the shortest way to connect pairs of changed ancillas. 
%
Moreover, the structure of their lattice provides an intrinsic advantage during the decoding. Indeed, it was explained above that a set of $\M{Z}$ ($\M{X}$) operators which forms a close loop on the edges of the lattice (including the link which is missing in each boundary plaquette) is contained in the stabilizer. Hence, if the decoder chooses a wrong error chain but the resulting correction operator applied after the original error closes a loop, the error is actually corrected. As a consequence, surface codes can also simply correct a large number of errors with weight larger than $t = \lfloor(d-1)/2 \rfloor$. 
For instance, it will be shown that a surface code with distance $d = 3$ will be able to correct, besides all errors of weight one, many errors with weight $w = 2$ and a good amount of errors with weight $w = 3, 4, 5, 6$.  

In general, a decoding error occurs every time that the actual error plus the correction operator realize a logical operator, so the whole chain operator crosses the lattice from boundary to boundary. For instance, consider a two qubits error $\M{Z}_2\M{Z}_3$ for the code in Fig.~\ref{Fig:surface}. In this scenario, the ancilla qubit $A_1$ is the only one that changes its state, hence the decoder locates an error $\M{Z}$ on data qubit $D_1$. It can be easily noticed that the whole operator applied to the lattice represents the logical $\M{Z}_L$.


\subsection{Asymptotic Performance Analysis} 

\begin{table*}[t]
    \centering
    \setlength{\tabcolsep}{3pt}
    \caption{Percentage of non-correctable error patterns per error class of surface codes using \acl{MWPM}.}
    \label{tab:Err}
    \small
    \begin{tabular}{lcccccccccccc}
        \toprule
        \rowcolor[gray]{.95}
        \textbf{Code} & $1-\beta_2$ & $1-\beta^{(Z)}_2$ & $\M{X}\M{X}$ & $\M{X}\M{Z}$ & $\M{X}\M{Y}$ & $\M{Z}\M{Z}$ & $\M{Z}\M{Y}$ & $\M{Y}\M{Y}$ & & & &\\
        \midrule
        $[[13,1,3]]$ & $0.24$ & $0.27$ & $0.27$ & $0$ & $0.28$ & $0.27$ & $0.26$ & $0.51$ \\
        $[[23,1,3/5]]$ & $0.07$ & $0$ & $0.16$ & $0$ & $0.16$ & $0$ & $0$ & $0.15$ \\
        \midrule 
        \rowcolor[gray]{.95}
        & $1-\beta_3$ & $1-\beta^{(Z)}_3$ & $\M{X}\M{X}\M{X}$ & $\M{X}\M{X}\M{Z}$ & $\M{X}\M{X}\M{Y}$ & $\M{X}\M{Z}\M{Z}$ & $\M{X}\M{Z}\M{Y}$ & $\M{X}\M{Y}\M{Y}$ & $\M{Z}\M{Z}\M{Z}$ & $\M{Z}\M{Z}\M{Y}$ & $\M{Z}\M{Y}\M{Y}$ & $\M{Y}\M{Y}\M{Y}$\\
        \midrule
        $[[13,1,3]]$ & $0.52$ & $0.53$ & $0.52$ & $0.27$ & $0.53$ & $0.27$ & $0.45$ & $0.67$ & $0.53$ & $0.53$ & $0.68$ & $0.78$ \\
        $[[23,1,3/5]]$ & $0.20$ & $0.08$ & $0.39$ & $0.15$ & $0.38$ & $0$ & $0.15$ & $0.39$ & $0.08$ & $0.08$ & $0.22$ & $0.45$ \\
        $[[41,1,5]]$ & $0.014$ & $0.024$ & $0.023$ & $0$ & $0.023$ & $0$ & $0$ & $0.023$ & $0.024$ & $0.024$ & $0.024$ & $0.046$ \\
        \bottomrule
    \end{tabular}
\end{table*}

In general, a code with distance $d$ can correct all patterns of up to $t = \lfloor(d-1)/2 \rfloor$ errors. For a decoder correcting only these patterns (bounded distance decoder), we can approximate the logical error probability with the asymptotic error probability given by
\begin{align}
\label{eq:error_probT}
\rho_\mathrm{L} 
&\simeq \binom{n}{t+1}\rho^{t+1}
\end{align}
which holds for $\rho \ll 1$.
From \eqref{eq:error_probT} we observe that the slope in log-log plot of the logical error probability is $t+1$.
Starting from \eqref{eq:PeAsym}, the asymptotic slope analysis can be extended to asymmetric codes correcting up to $e_\mathrm{g}$ generic errors and $e_\mathrm{Z}$ Pauli $\M{Z}$ errors, giving
\begin{align}
\label{eq:PeAsymApprox}
\rho_\mathrm{L} &\simeq 
\begin{dcases}
(1 - \alpha_{e_\mathrm{g}+1}) \,\binom{n}{e_\mathrm{g}+1}\,\rho^{e_\mathrm{g}+1} & 1 \le A < \infty\\
\binom{n}{e_\mathrm{g} + e_\mathrm{Z} +1}\,\rho^{e_\mathrm{g} + e_\mathrm{Z}+1} & A \to \infty\,.\\
\end{dcases}
\end{align}
We observe that in this case the asymptotic slope is $e_\mathrm{g} + 1$ for all $A < \infty$. On the other hand, considering a phase-flip channel ($A \to \infty$) the slope becomes $e_\mathrm{g} + e_\mathrm{Z} + 1$.

However, as previously observed, for surface codes it is easy to implement decoders which always try to correct the error (complete decoders), therefore allowing to recover a large number of errors of weight $j > t$. In order to account for this capability, we rewrite the error probability in \eqref{eq:error_probT} taking into account that a percentage $\beta_j$ of errors of weight $j$ can be corrected. Since the code guarantees that all error patterns with less than $t+1$ errors are corrected, we have $\beta_{j} = 1$ for  $j \leq t$. Over a depolarizing channel, the performance of complete decoders is then
\begin{align}
\label{eq:error_probWithBeta}
\rho_\mathrm{L} 
&=  \sum_{j = t+1}^{n} (1-\beta_j) \binom{n}{j} \rho^j (1-\rho)^{n-j} 
\end{align}
which, for $\rho \ll 1$, can be approximated as
\begin{align}
\label{eq:error_probWithBetaApprox}
\rho_\mathrm{L} 
&\simeq \left(1-\beta_{t+1}\right) \binom{n}{t+1}\rho^{t+1} \,.
\end{align}
Note that \eqref{eq:error_probWithBetaApprox} differs from \eqref{eq:error_probT} in that the latter assumes that all errors with weight greater than $t$ are not correctable. 
The asymptotic slope in loglog domain remains $t+1$, while an offset at low $\rho$, depending on $(1-\beta_{t+1})$, appears when compared to \eqref{eq:error_probT}.
Over phase-flip channels ($A \to \infty$) the error patterns have only Pauli $\M{Z}$ errors, leading in general to different values of the $\beta_j$ for the same code over a depolarizing channel.
For this reason, we adopt the notation $\beta_j$ when considering depolarizing channels and $\beta^{(Z)}_j$ for phase-flip ones. Hence, the error probability can be written as
\begin{align}
\label{eq:error_probWithBetaPhaseFlip}
\rho_\mathrm{L} 
&=  \sum_{j = t_\mathrm{Z}+1}^{n} \left(1-\beta^{(Z)}_j\right) \binom{n}{j} \pz^j (1-\pz)^{n-j}\,. 
\end{align}
The direct approach to compute the $\beta_j$ and $\beta^{(Z)}_j$ coefficients for a given code is by exhaustive search. 
For example, the error patterns of weight two can be of type $\M{X}\M{X}$, $\M{X}\M{Z}$, $\M{X}\M{Y}$, $\M{Z}\M{Z}$, $\M{Z}\M{Y}$, and $\M{Y}\M{Y}$. 
In general, the total number of error classes of weight $j$ are $\binom{j + 2}{2}$.
Limiting the search to phase-flip channels, the only class of errors is that including just  $\M{Z}$ operators. 

In Tab.~\ref{tab:Err} we report the percentage of non-correctable errors for each error class, which we have evaluated by exhaustive search with a \ac{MWPM} decoder. In addition, we computed the value of $1-\beta_j$ and $1-\beta^{(Z)}_j$ by weighting the percentages of non-correctable errors. The total number of different error combinations within the error class $i$ with error pattern of weight $j$ is
\begin{align}
c^{(j)}_i = \binom{n}{n_\mathrm{x}}\binom{n-n_\mathrm{x}}{n_\mathrm{z}}\binom{n-n_\mathrm{x}-n_\mathrm{z}}{n_\mathrm{y}}
\end{align}
where $n_\mathrm{x}$, $n_\mathrm{z}$, and $n_\mathrm{y}$ are the number of errors $\M{X}$, $\M{Z}$, and $\M{Y}$ in the class $i$, respectively, and $j = n_\mathrm{x} + n_\mathrm{z} + n_\mathrm{y}$.
For example, the error class $\M{X}\M{X}\M{Z}$ has $n_\mathrm{x} = 2$, $n_\mathrm{z} = 1$, and $n_\mathrm{y} = 0$.
Defining as $p_i$ the percentage of non-correctable errors in the class $i$, we have that
\begin{align}
\label{eq:computeBeta}
    1-\beta_j = \frac{\displaystyle \sum_{i=1}^{K_j} c^{(j)}_i \, p_i}{\displaystyle\sum_{i=1}^{K_j} c^{(j)}_i} \quad, \quad  K_j=\binom{j + 2}{2}\,.
\end{align}
On the other hand, the value of $1-\beta^{(Z)}_j$ is equivalent to $p_i$ of the class with $j$ Pauli $\M{Z}$.
From Tab.~\ref{tab:Err}, we observe that, as anticipated, surface codes can correct a large amount of errors above the nominal error correction capability. As an example, the $[[13,1,3]]$ is able to correct about $75\%$ and $50\%$ of the error patterns of weight $2$ and $3$, respectively.

Note that, increasing the distance of the code, the exhaustive search for deriving the $\beta_j$ could become infeasible. However, as can be observed in Tab.~\ref{tab:Err}, the value of $\beta^{(Z)}_j$ is generally close to $\beta_j$. Hence, it is possible to obtain a good estimation of $\beta_j$ taking into account just $\M{Z}$ errors, which require to consider only $\binom{n}{j}$ patterns.

\subsection{Code-Effective Threshold}

We define the code-effective threshold $\rho_\mathrm{thr}$ as  
\begin{align}
\label{eq:rhothr}
    \rho_{thr}= \max\left\{\rho: \rho_L \leq   10^{-\gamma} \, \rho\right\} \,.
\end{align}
This threshold is the value of $\rho$ such that a \ac{QECC} gives a performance boost in terms of logical error rate of $10^{\gamma}$, with $\gamma \ge 0$, compared to the uncoded case. 
In \cite{AzaLipMah:22}, the authors define a pseudo-threshold which is equivalent to the code-effective threshold when $\gamma = 0$. 
Despite the fact that a quantum code could improve the performance for some values of $\rho$, if the performance boost is small, the deployment of coding and decoding schemes may not be worth it. For this reason, we generalize this definition aiming to target a performance boost requirement.

In general, to compute this threshold it is necessary to proceed by Monte Carlo simulations in order to find the intersection with the shifted uncoded performance $\rho_\mathrm{L} = 10^{-\gamma} \, \rho$. However, if the code performance are described by an analytical expressions, as in the previous section, it is possible to derive this value without simulations. For surface codes, we can use \eqref{eq:error_probWithBeta} over depolarizing channels and \eqref{eq:error_probWithBetaPhaseFlip} over phase-flip channels. In this case we will need an exhaustive search to compute the $\beta_j$ and $\beta_j^{(Z)}$. To reduce the complexity, we can exploit the asymptotic behavior to approximate the code-effective threshold as
\begin{align}
\label{eq:codeEffThr}
\tilde{\rho}_\mathrm{thr} &= 
\sqrt[t]{\frac{1}{10^\gamma\,(1-\beta_{t+1}) \, \binom{n}{t+1}}}\,.
\end{align}

\section{Numerical Results}\label{sec:NumRes}

\begin{figure}[t]
	\centering
	\resizebox{0.49\textwidth}{!}{
%
%
\definecolor{mycolor1}{rgb}{0.92900,0.69400,0.12500}%
\definecolor{mycolor4}{rgb}{0.07843,0.35294,0.19608}%
\definecolor{mycolor6}{rgb}{0.12157,0.38039,0.54510}%

\definecolor{darkGreen}{rgb}{0.07843,0.35294,0.19608}%
\definecolor{darkRed}{rgb}{0.57255,0.16863,0.12941}%
\definecolor{darkBlue}{rgb}{0.12157,0.38039,0.54510}%
\definecolor{redRose}{HTML}{d71c36}
\definecolor{herbalGreen}{HTML}{196f3d}
\begin{tikzpicture}

\begin{axis}[%
name = match,
width=4.5in,
height=3.5in,
at={(0in,0in)},
scale only axis,
xmode=log,
xmin=0.001,
xmax=0.2,
xminorticks=true,
xlabel={$\rho$},
xlabel style={font=\color{white!15!black}, font = \Large},
ymode=log,
ymin=1e-05,
ymax=1,
yminorticks=true,
ylabel={$\rho_\mathrm{L}$},
ylabel style={font=\color{white!15!black}, font = \Large},
axis background/.style={fill=white},
tick label style={black, semithick, font=\Large},
xmajorgrids,
ymajorgrids,
legend style={at={(0.03,0.97)}, anchor=north west, legend cell align=left, align=left, draw=white!15!black}
]
\addplot [only marks, color=darkBlue, line width=1.3pt, mark=o, mark size = 3.0pt, mark options={solid, darkBlue}]
 table[row sep=crcr, y error plus index=2, y error minus index=3]{%
0.001	0	0	0\\
0.006	0.00066	3.97946494443663e-05	3.97946494443663e-05\\
0.01021429	0.0019	0.00019085559756004	0.00019085559756004\\
0.01942857	0.0066	0.000354874765966813	0.000354874765966813\\
0.02864286	0.01402	0.000515286937382271	0.000515286937382271\\
0.03785714	0.0228	0.000654184464260655	0.000654184464260655\\
0.04707143	0.03548	0.000810752875836281	0.000810752875836281\\
0.05628571	0.04958	0.000951375839355194	0.000951375839355194\\
0.07471429	0.08158	0.00119964732598743	0.00119964732598743\\
0.10235714	0.13358	0.0014909944520986	0.0014909944520986\\
0.13	0.194538	0.00173486555721406	0.00173486555721406\\
0.1574	0.2539	0.00190752809843525	0.00190752809843525\\
};\label{plot:3x3SimFig2}

\addplot [only marks, color=darkGreen, line width=1.3pt, mark=square,mark size = 2.5pt, mark options={solid, darkGreen}]
 table[row sep=crcr, y error plus index=2, y error minus index=3]{%
0.001	0	0	0\\
0.006	2.5e-05	7.90090987950122e-06	7.90090987950122e-06\\
0.01191143	0.000190200000000001	6.04372958579691e-05	6.04372958579691e-05\\
0.02334286	0.00134	0.000160325388242786	0.000160325388242786\\
0.03451429	0.00422	0.000284105079596969	0.000284105079596969\\
0.04568571	0.01004	0.000436934918342767	0.000436934918342767\\
0.05685714	0.01862	0.000592447011879375	0.000592447011879375\\
0.0792	0.04288	0.000887874652169325	0.000887874652169325\\
0.10154286	0.07914	0.00118313935551278	0.00118313935551278\\
0.12388571	0.12246	0.00143671842671666	0.00143671842671666\\
0.1574	0.1969	0.00174280478858649	0.00174280478858649\\
};\label{plot:3x5SimFig2}

\addplot [color=brightBlue, line width=1.3pt]
  table[row sep=crcr]{%
0.001	1.86627311743459e-05\\
0.01021429	0.00189145955855941\\
0.01942857	0.0066388598947722\\
0.02864286	0.0139809935654848\\
0.03785714	0.0236369133710378\\
0.04707143	0.0353287502302211\\
0.05628571	0.0487850769602719\\
0.0655	0.0637438822701075\\
0.07471429	0.0799548394343301\\
0.08392857	0.0971812582023263\\
0.09314286	0.115201658401447\\
0.10235714	0.133810788305131\\
0.11157143	0.152820609008863\\
0.12078571	0.172060673708408\\
0.13	0.191378572425128\\
0.139	0.210193439159267\\
0.1482	0.229257252790112\\
0.1574	0.248051791686608\\
};\label{plot:3x3AppFig2}

\addplot [color=graphGreen, line width=1.3pt]
  table[row sep=crcr]{%
0.001	1.34411541008461e-07\\
0.01217143	0.000230358832087552\\
0.02334286	0.00153898860213475\\
0.03451429	0.00469551510966459\\
0.04568571	0.0102461639883242\\
0.05685714	0.0185292018312047\\
0.06802857	0.0296998814810278\\
0.0792	0.0437622906749204\\
0.09037143	0.0606046972289827\\
0.10154286	0.0800351034651324\\
0.11271429	0.101814132147396\\
0.12388571	0.125683046745941\\
0.13505714	0.151385677150334\\
0.14622857	0.17868337334861\\
0.1574	0.207363578033399\\
};\label{plot:3x5AppFig2}

\addplot [color=brightBlue,dashed, line width=1.5pt]
  table[row sep=crcr]{%
0.001	2.067e-05\\
0.01021429	0.00215653665661875\\
0.01942857	0.00780229109750208\\
0.02864286	0.0169579455770083\\
0.03785714	0.0296234802224083\\
0.04707143	0.0457989225248021\\
0.05628571	0.0654842373747188\\
0.0655	0.0886794675\\
0.07471429	0.115384593441319\\
0.08392857	0.145599580502602\\
0.09314286	0.179324494266808\\
0.10235714	0.216559261532608\\
0.11157143	0.257303963119702\\
0.12078571	0.301558510590019\\
0.13	0.349323\\
0.139	0.39936507\\
0.1482	0.4539801708\\
0.1574	0.5120942892\\
};\label{plot:3x3AsympAppFig2}

\addplot [color=graphGreen, dashed, line width=1.5pt]
  table[row sep=crcr]{%
0.001	1.35076271186441e-07\\
0.01217143	0.000243558830764628\\
0.02334286	0.00171807160160884\\
0.03451429	0.00555361660582365\\
0.04568571	0.0128801286025557\\
0.05685714	0.0248275630824709\\
0.06802857	0.0425258584353582\\
0.0792	0.0671049578782373\\
0.09037143	0.0996948046281281\\
0.10154286	0.14142534190205\\
0.11271429	0.193426512917024\\
0.12388571	0.256828198696886\\
0.13505714	0.332760455122717\\
0.14622857	0.4223531739292\\
0.1574	0.526736298333356\\
};\label{plot:3x5AsympAppFig2}

\end{axis}

\node[draw,fill=white,inner sep=1.5pt,below right=0.5em] at (match.north west){
  {\renewcommand{\arraystretch}{1.2}
    \begin{tabular}{lcc}
        & \textbf{$d = 3$} & \textbf{$d = 3/5$}\\
        & \textbf{$A = 1$} & \textbf{$A \to \infty$}\\
        \textbf{Simulation} & \ref{plot:3x3SimFig2} & \ref{plot:3x5SimFig2}\\
        \textbf{Asympt. Approx.} & \ref{plot:3x3AsympAppFig2} & \ref{plot:3x5AsympAppFig2}\\
        \textbf{Approximation} & \ref{plot:3x3AppFig2} & \ref{plot:3x5AppFig2}\\
  \end{tabular}
}};

\end{tikzpicture}%
	}
	\caption{Logical error rate vs.\ physical error rate, $[[13,1,3]]$ surface code over a depolarizing channel, $[[23,1,3/5]]$ surface code over a phase flip channel. Comparison between simulation and theoretical approximations.
		\label{Fig:plot_3x3_matching}}
\end{figure}
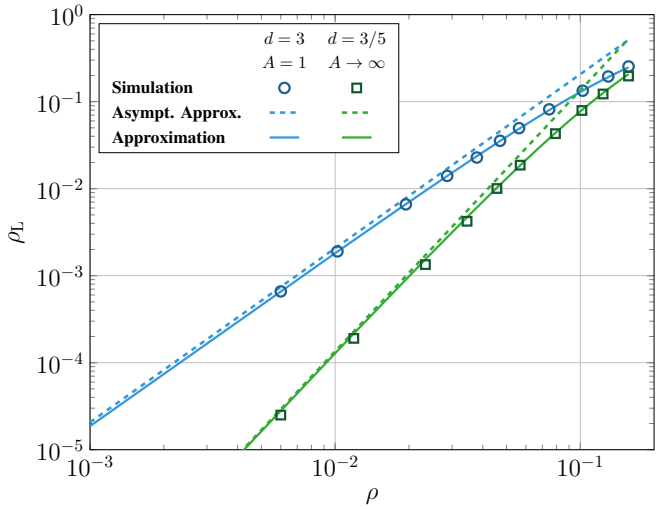

\emph{1) Asymptotic approximations and matching.}
In Fig.~\ref{Fig:plot_3x3_matching} we verify the asymptotic approximations \eqref{eq:error_probWithBetaApprox} and the corresponding approximation for phase-flip channels that can be derived from \eqref{eq:error_probWithBetaPhaseFlip}. To this aim, we compare the analytical formulas with simulations for the $[[13,1,3]]$ surface code over a depolarizing channel, and for the $[[23,1,3/5]]$ surface code (described later in this section) over a phase-flip channel. For both the codes we consider the \ac{MWPM} decoder.
As expected, the asymptotic approximations are tights for $\rho \ll 1$. Hence, it is possible to use this tool to compute logical error rates not achievable by Monte Carlo simulations.
Moreover, we report two approximations aiming to match the simulated curves. For the $[[13,1,3]]$ surface code over depolarizing channel we plot \eqref{eq:error_probWithBeta} using $(\beta_2, \beta_3, \beta_4, \beta_5, \beta_6) = (0.76, 0.48, 0.48, 0.46, 0.5)$ where $(\beta_2, \beta_3)$ are computed by exhaustive search according to \eqref{eq:computeBeta}, while $(\beta_4, \beta_5, \beta_6)$ are approximated using $\left(\beta^{(Z)}_4, \beta^{(Z)}_5, \beta^{(Z)}_6\right)$. The $\beta_j$ with $j>6$ are set to zero.
Similarly, for the $[[23,1,3/5]]$ surface code over phase-flip channel we plot \eqref{eq:error_probWithBetaPhaseFlip} using $\left(\beta^{(Z)}_3, \beta^{(Z)}_4, \beta^{(Z)}_5, \beta^{(Z)}_6, \beta^{(Z)}_7\right) = (0.92, 0.76, 0.59, 0.52, 0.49)$. The $\beta^{(Z)}_j$ with $j>7$ are set to zero.
For both codes, from Fig.~\ref{Fig:plot_3x3_matching} we observe that the approximations are tight also for large $\rho$. 

\begin{figure}[t]
	\centering
	\resizebox{0.49\textwidth}{!}{
	    \input{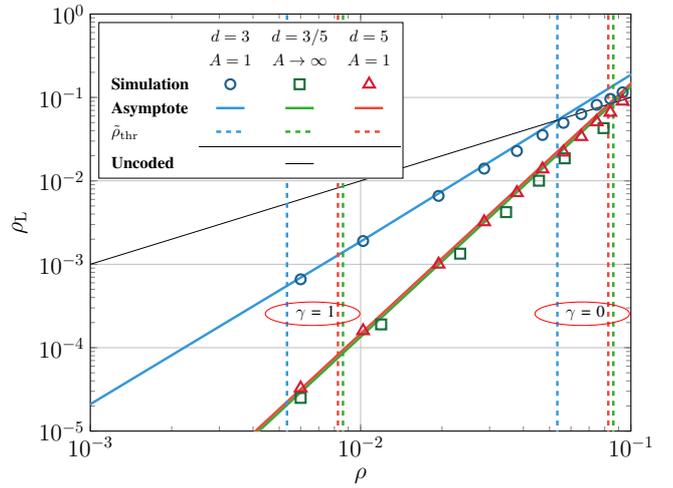}
	}
	\caption{Logical error rate and code-effective threshold, $[[13,1,3]]$ and $[[41,1,5]]$ surface codes over a depolarizing channel, $[[23,1,3/5]]$ surface code over a phase flip channel, $\gamma \in \{0,1\}$.
		\label{Fig:plot_plot_thre}}
\end{figure}

\emph{2) Code effective threshold.} Fig.~\ref{Fig:plot_plot_thre} shows that the code effective  threshold, obtained with \eqref{eq:codeEffThr}, increases from $\tilde{\rho}_\mathrm{thr} = 0.0534$ to $\tilde{\rho}_\mathrm{thr} = 0.0824$ when we pass from the $[[13,1,3]]$ to the $[[41,1,5]]$ surface code. As expected, the threshold value increases with the error correction capability of the code. 
For the $[[23,1,3/5]]$ asymmetric surface code on a phase flip channel, we can approximate the performance according to \eqref{eq:PeAsymApprox} letting $A \to \infty$, which gives an error probability as in \eqref{eq:error_probWithBetaApprox} with the related substitutions. Since the $[[23,1,3/5]]$ is able to correct $e_\mathrm{g} = 1$ and $e_\mathrm{Z} = 1$, and in a phase flip channel only $\M{Z}$ errors can occur, $\beta_{t+1}$ is the fraction of $\M{Z}\M{Z}\M{Z}$ errors corrected. Hence, it is possible to obtain $\tilde{\rho}_\mathrm{thr}$ by substituting $ t = e_\mathrm{g}  + e_\mathrm{Z}$ in \eqref{eq:codeEffThr}.
The resulting threshold, $\tilde{\rho}_\mathrm{thr} = 0.086$,  is higher than the previous ones. This is due to the fact that this code has distance $d_\mathrm{z} = 5$, thus on a phase flip channel it works better than the $[[13,1,3]]$ code (which has smaller distance), and also of the $[[41,1,5]]$ code (which uses a larger number of qubits). Thus, we have shown that this figure of merit can be simply used to compare the performances of symmetric topological codes over depolarizing channels, and asymmetric codes over channels with only one kind of error.

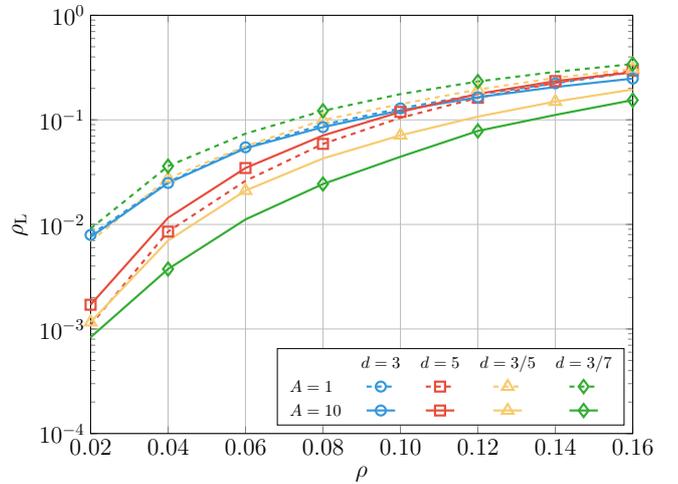
\begin{figure}[t]
	\centering
	\resizebox{0.49\textwidth}{!}{
%
%
\definecolor{yellowPaglia}{HTML}{f7dc6f}%
\definecolor{yellowArancio}{HTML}{f7cb6f}%
\definecolor{darkGreen}{rgb}{0.07843,0.35294,0.19608}%
\definecolor{darkRed}{rgb}{0.57255,0.16863,0.12941}%
\definecolor{darkBlue}{rgb}{0.12157,0.38039,0.54510}%
\definecolor{redRose}{HTML}{d71c36}
\definecolor{herbalGreen}{HTML}{196f3d}
\begin{tikzpicture}

\begin{axis}[%
name = asymmetrical,
width=4.5in,
height=3.5in,
at={(0in,0in)},
scale only axis,
xmin=0.02,
xmax=0.16,
xtick = {0.02, 0.04, 0.06, 0.08, 0.10, 0.12, 0.14, 0.16},
xticklabels = {$0.02$, $0.04$, $0.06$, $0.08$, $0.10$, $0.12$, $0.14$, $0.16$},
xminorticks=true,
xlabel={$\rho$},
xlabel style={font=\color{white!15!black}, font = \Large},
ymode=log,
ymin=0.0001,
ymax=1,
yminorticks=true,
ylabel={$\rho_\mathrm{L}$},
ylabel style={font=\color{white!15!black}, font = \Large},
axis background/.style={fill=white},
tick label style={black, semithick, font=\Large},
xmajorgrids,
ymajorgrids,
legend style={at={(0.03,0.97)}, anchor=north west, font = \large,  legend cell align=left, align=left, draw=white!15!black}
]
\addplot [color=brightBlue, dashed, line width=1.3pt, mark=o,mark size = 3.0pt, mark options={solid, brightBlue}, mark repeat = 2]
 table[row sep=crcr, y error plus index=2, y error minus index=3]{%
0.02	0.00792824554319447	0.000293820529916878	0.000293820529916878\\
0.04	0.0254252668814676	0.000521509690989304	0.000521509690989304\\
0.06	0.0546384094894819	0.00075295678321442	0.00075295678321442\\
0.08	0.0898237324030709	0.00094728432595223	0.00094728432595223\\
0.1	0.12871040550601	0.00110945615070488	0.00110945615070488\\
0.12	0.174759075361729	0.00125814963742883	0.00125814963742883\\
0.14	0.223392308142566	0.00137992982300659	0.00137992982300659\\
0.16	0.286590462555823	0.00149803641802949	0.00149803641802949\\
}; \label{3x3_A1}

\addplot [color=brightRed, dashed, line width=1.3pt, mark=square,mark size = 3.0pt, mark options={solid, brightRed}, mark repeat = 2, mark phase = 2]
 table[row sep=crcr, y error plus index=2, y error minus index=3]{%
0.02	0.00109143033234475	0.000109391336472438	0.000109391336472438\\
0.04	0.00853300681560655	0.00030472790492923	0.00030472790492923\\
0.06	0.026101340218729	0.000528214522127598	0.000528214522127598\\
0.08	0.0588977643145722	0.000779991476095945	0.000779991476095945\\
0.1	0.10452243345525	0.00101357096352137	0.00101357096352137\\
0.12	0.162476157555164	0.00122212452305616	0.00122212452305616\\
0.14	0.225674563654123	0.00138492137372153	0.00138492137372153\\
0.16	0.295491641027319	0.00151160292744711	0.00151160292744711\\
};\label{5x5_A1}

\addplot [color=yellowArancio, dashed, line width=1.3pt, mark=triangle, mark size = 4.0pt, mark options={solid, yellowArancio}, mark repeat = 2, mark phase = 2]
 table[row sep=crcr, y error plus index=2, y error minus index=3]{%
0.02	0.00687689913555826	0.000273791473122816	0.000273791473122816\\
0.04	0.027574685420523	0.000542507142679096	0.000542507142679096\\
0.06	0.0560007953224596	0.000761736841535796	0.000761736841535796\\
0.08	0.0997662399573169	0.000992867967076854	0.000992867967076854\\
0.1	0.141266543385522	0.00115390724531445	0.00115390724531445\\
0.12	0.194649906510395	0.00131172110327426	0.00131172110327426\\
0.14	0.25170271616279	0.00143781524689981	0.00143781524689981\\
0.16	0.306572900450526	0.00152752854827228	0.00152752854827228\\
}; \label{3/5_A1}

\addplot [color=graphGreen, dashed, line width=1.3pt, mark=diamond, mark size = 4.0pt, mark options={solid, graphGreen}, mark repeat = 2, mark phase = 2]
 table[row sep=crcr, y error plus index=2, y error minus index=3]{%
0.02	0.00911534101862667	0.000314861863848747	0.000314861863848747\\
0.04	0.0360788721819864	0.0006178304025481	0.0006178304025481\\
0.06	0.0737730691208126	0.000866023500168993	0.000866023500168993\\
0.08	0.12196414357886	0.00108416219749125	0.00108416219749125\\
0.1	0.176326475221536	0.00126257843009761	0.00126257843009761\\
0.12	0.232626137366157	0.00139976395301526	0.00139976395301526\\
0.14	0.287768666975438	0.00149987248972638	0.00149987248972638\\
0.16	0.341885248302104	0.00157149514311803	0.00157149514311803\\
};\label{3/7_A1}

\addplot [color=brightBlue, line width=1.3pt, mark=o,mark size = 3.0pt, mark options={solid, brightBlue}, mark repeat = 2, mark phase = 2]
 table[row sep=crcr, y error plus index=2, y error minus index=3]{%
0.02	0.00758799449837798	0.00028749583512195	0.00028749583512195\\
0.04	0.0248902135910529	0.000516134771994959	0.000516134771994959\\
0.06	0.0534937692102666	0.000745478954810004	0.000745478954810004\\
0.08	0.08522208676395	0.0009250303167045	0.0009250303167045\\
0.1	0.123119203480125	0.00108856715759518	0.00108856715759518\\
0.12	0.164475181918336	0.0012281514184534	0.0012281514184534\\
0.14	0.206099672222857	0.00134012000519408	0.00134012000519408\\
0.16	0.248244737674162	0.00143119995245051	0.00143119995245051\\
};\label{3x3_A10}

\addplot [color=brightRed, line width=1.3pt, mark=square,mark size = 3.0pt, mark options={solid, brightRed}, mark repeat = 2]
 table[row sep=crcr, y error plus index=2, y error minus index=3]{%
0.02	0.00170518041258779	0.000136689962869609	0.000136689962869609\\
0.04	0.0115661957264477	0.00035423470922439	0.00035423470922439\\
0.06	0.034633285720866	0.000605780176253471	0.000605780176253471\\
0.08	0.0708594217715302	0.000850083457406539	0.000850083457406539\\
0.1	0.119139743609413	0.00107325738989331	0.00107325738989331\\
0.12	0.176951655272903	0.00126433464013791	0.00126433464013791\\
0.14	0.234781348254313	0.00140425706935226	0.00140425706935226\\
0.16	0.284814461236901	0.00149524524361919	0.00149524524361919\\
};\label{5x5_A10}

\addplot [color=yellowArancio, line width=1.3pt, mark=triangle, mark size = 4.0pt, mark options={solid, yellowArancio}, mark repeat = 2]
 table[row sep=crcr, y error plus index=2, y error minus index=3]{%
0.02	0.00116208741900153	0.000112872711506499	0.000112872711506499\\
0.04	0.0070089954104542	0.000276390170712064	0.000276390170712064\\
0.06	0.0210535323002459	0.000475624268841848	0.000475624268841848\\
0.08	0.0427486056640962	0.000670187047218551	0.000670187047218551\\
0.1	0.071114566350366	0.000851495600693006	0.000851495600693006\\
0.12	0.107466513759787	0.00102605558567206	0.00102605558567206\\
0.14	0.149498707583118	0.00118134910669908	0.00118134910669908\\
0.16	0.195176697032629	0.00131306523354838	0.00131306523354838\\
}; \label{3/5_A10}

\addplot [color=graphGreen, line width=1.3pt, mark=diamond, mark size = 4.0pt, mark options={solid, graphGreen}, mark repeat = 2, mark phase = 2]
 table[row sep=crcr, y error plus index=2, y error minus index=3]{%
0.02	0.000833244767235761	9.55932736286426e-05	9.55932736286426e-05\\
0.04	0.00374299380398173	0.000202309975541947	0.000202309975541947\\
0.06	0.0111650075396755	0.000348107574186412	0.000348107574186412\\
0.08	0.0243431982339158	0.000510574834577511	0.000510574834577511\\
0.1	0.0443890941982264	0.00068233985578963	0.00068233985578963\\
0.12	0.0783367517435677	0.000890206915973192	0.000890206915973192\\
0.14	0.111483815630685	0.00104270302345711	0.00104270302345711\\
0.16	0.154422522839462	0.00119716517285554	0.00119716517285554\\
}; \label{3/7_A10}

\end{axis}

\node[draw,fill=white,inner sep=1.5pt,above left=0.5em] at (asymmetrical.south east){
  {\renewcommand{\arraystretch}{1.2}
    \begin{tabular}{lcccc}
        & $d = 3$ & $d = 5$ & $d = 3/5$ & $d = 3/7$\\
        $A=1$ &   \ref{3x3_A1} & \ref{5x5_A1} &  \ref{3/5_A1} & \ref{3/7_A1}
         \\
        $A=10$ & \ref{3x3_A10} & \ref{5x5_A10} & \ref{3/5_A10} & \ref{3/7_A10} \\
        
      
  \end{tabular}
}};

\end{tikzpicture}%
	}
	\caption{Logical error rate, $[[13,1,3]]$ and $[[41,1,5]]$ symmetric surface codes, $[[23,1,3/5]]$ and $[[33,1,3/7]]$ asymmetric surface codes. Depolarizing $(A=1)$ and polarizing $(A=10)$ channels.
		\label{Fig:plot_asymmetrical}}
\end{figure}

\emph{3) Performance over symmetric and asymmetric channels.} In Fig.~\ref{Fig:plot_asymmetrical} we investigate the advantage that a rectangular surface code is able to achieve over an asymmetric channel. Such a code has the horizontal and the vertical dimensions of different length. In particular, since $\M{Z}$ errors often happen more frequently in hardware implementations  \cite{IofMez:07,ChiVal:20a,AzaLipMah:22}, we have considered the $[[23,1,3/5]]$ surface code, in which the horizontal direction is two-qubit longer than the vertical one, making the logical $\M{Z}_L$ operator a chain of 5 qubits (not only 3 like in the $[[13,1,3]]$ code). A code of this kind requires a lattice of 23 qubits but it has a distance $d_{Z} = 5$, so it is able to correct $\M{Z}$ errors of weight $w = 2$. 
In particular, Fig.~\ref{Fig:plot_asymmetrical} shows that the $[[23,1,3/5]]$ code has performance similar to the $[[13,1,3]]$ code over a symmetric channel, while the $[[41,1,5]]$ code has a lower logical error rate. Indeed, in the case in which errors of different kind occur with the same probability, there is no advantage in making a direction of the lattice longer than the other, since, in this case, the code can still correct only $\M{X}$ error of weight $w = 1$. Moreover, having a lattice with 23 qubits instead of 13 determines, on average, a larger number of data qubits in error. However, if we consider a polarizing channel $(A = 10)$, in which $\M{Z}$'s are more probable, the performance of the $[[41,1,5]]$ code gets worse, while the $[[23,1,3/5]]$ can exploit its asymmetry. In fact, this code is able to correct $\M{Z}$ errors of weight $w = 2$ like the $[[41,1,5]]$, but using about half of its qubits, resulting in a lower logical error rate. These considerations can be made even more evident if we consider the $[[33,1,3/7]]$ surface code. This code has the worst performance over a depolarizing channel, since it uses a higher number of qubits, but it still has a distance $d_{\mathrm{x}} = 3$. Contrarily, it has the best error correction capability over a channel with $A = 10$,  since it is able to correct all $\M{Z}$ errors of weight up to $w = 3$.  

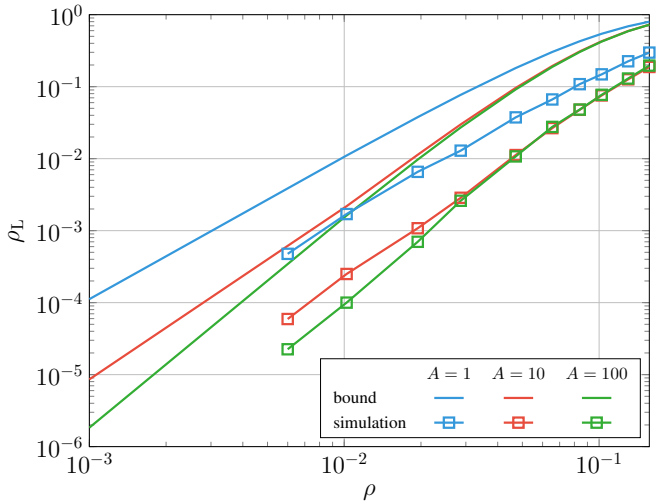
\begin{figure}[t]
	\centering
	\resizebox{0.49\textwidth}{!}{
%
%
\definecolor{mycolor1}{rgb}{0.00000,0.44700,0.74100}%
\definecolor{mycolor2}{rgb}{0.85000,0.32500,0.09800}%
\definecolor{mycolor3}{rgb}{0.92900,0.69400,0.12500}%
\definecolor{mycolor4}{rgb}{0.49400,0.18400,0.55600}%
\definecolor{mycolor5}{rgb}{0.46600,0.67400,0.18800}%
\definecolor{mycolor6}{rgb}{0.30100,0.74500,0.93300}%
\definecolor{mycolor7}{rgb}{0.63500,0.07800,0.18400}%
\definecolor{mycolor8}{rgb}{0.13500,0.03800,0.58400}%

\definecolor{yellowPaglia}{HTML}{f7dc6f}%
\definecolor{yellowArancio}{HTML}{f7cb6f}%
\definecolor{darkGreen}{rgb}{0.07843,0.35294,0.19608}%
\definecolor{darkRed}{rgb}{0.57255,0.16863,0.12941}%
\definecolor{darkBlue}{rgb}{0.12157,0.38039,0.54510}%
\definecolor{redRose}{HTML}{d71c36}
\definecolor{herbalGreen}{HTML}{196f3d}
\begin{tikzpicture}

\begin{axis}[%
name = asymmetrical_3x5,
width=4.5in,
height=3.5in,
at={(0in,0in)},
scale only axis,
xmode=log,
xmin=0.001,
xmax=0.1574,
xminorticks=true,
xlabel={$\rho$},
xlabel style={font=\color{white!15!black}, font = \Large},
ymode=log,
ymin=1e-06,
ymax=1,
yminorticks=true,
ylabel={$\rho_\mathrm{L}$},
ylabel style={font=\color{white!15!black}, font = \Large},
axis background/.style={fill=white},
tick label style={black, semithick, font=\Large},
xmajorgrids,
ymajorgrids,
legend style={at={(0.58,0.41)}, anchor=north west, font = \large, legend cell align=left, align=left, draw=white!15!black}
]
\addplot [color=brightBlue, line width=1.3pt,  mark options={solid, brightBlue}] 
  table[row sep=crcr]{%
0.001	0.000111832860356209\\
0.01021429	0.0110596449683333\\
0.01942857	0.0377315884416656\\
0.02864286	0.0769962716875318\\
0.04707143	0.181430464778123\\
0.0655	0.303416812452847\\
0.08392857	0.427381832950258\\
0.10235714	0.543141854244788\\
0.13	0.689945229866124\\
0.1574	0.79941224838222\\
};\label{3x5_A1}

\addplot [color=brightRed, line width=1.3pt,  mark options={solid, brightRed}] 
  table[row sep=crcr]{%
0.001	8.59762327354296e-06\\
0.01021429	0.00218535143658327\\
0.01942857	0.0113258783762855\\
0.02864286	0.029883150274047\\
0.04707143	0.0961733054831202\\
0.0655	0.193669813650193\\
0.08392857	0.308565867393847\\
0.10235714	0.427598896998312\\
0.13	0.593048363012162\\
0.1574	0.726675841830384\\
};\label{3x5_A10}

\addplot [color=graphGreen, line width=1.3pt, mark options={solid, graphGreen}] 
  table[row sep=crcr]{%
0.001	1.83521477603588e-06\\
0.01021429	0.00162357325911333\\
0.01942857	0.00971229989973277\\
0.02864286	0.0271075087276018\\
0.04707143	0.0915233011054641\\
0.0655	0.188162329531215\\
0.08392857	0.303119115281227\\
0.10235714	0.422801989970082\\
0.13	0.589651431960864\\
0.1574	0.724589064723373\\
};\label{3x5_A100}

\addplot [color=brightBlue, line width=1.3pt, mark=square,mark size = 3.0pt, mark options={solid, brightBlue}]
 table[row sep=crcr, y error plus index=2, y error minus index=3]{%
0.006	0.000476	4.04453265754945e-05	4.04453265754945e-05\\
0.01021429	0.0017	8.0744150103893e-05	8.0744150103893e-05\\
0.01942857	0.00656000000000001	0.000500355654732112	0.000500355654732112\\
0.02864286	0.01292	0.000699944337370909	0.000699944337370909\\
0.04707143	0.03744	0.00117662452537009	0.00117662452537009\\
0.0655	0.0664	0.00154319402300553	0.00154319402300553\\
0.08392857	0.1083	0.00192610377751564	0.00192610377751564\\
0.10235714	0.14828	0.00220265146723343	0.00220265146723343\\
0.13	0.22522	0.00258909954939272	0.00258909954939272\\
0.1574	0.2973	0.00283294734637974	0.00283294734637974\\
};\label{3x5_A1_sim}

\addplot [color=brightRed, line width=1.3pt, mark=square,mark size = 3.0pt, mark options={solid, brightRed}]
 table[row sep=crcr, y error plus index=2, y error minus index=3]{%
0.006	5.92e-05	1.5080094713727e-05	1.5080094713727e-05\\
0.01021429	0.00025	3.09864470373743e-05	3.09864470373743e-05\\
0.01942857	0.00108	0.000203579153101687	0.000203579153101687\\
0.02864286	0.00286	0.00033099174081297	0.00033099174081297\\
0.04707143	0.01122	0.000652833356397787	0.000652833356397787\\
0.0655	0.02654	0.000996243695395057	0.000996243695395057\\
0.08392857	0.04774	0.0013215239848139	0.0013215239848139\\
0.10235714	0.07554	0.00163790468034449	0.00163790468034449\\
0.13	0.12632	0.00205905679959578	0.00205905679959578\\
0.1574	0.18774	0.00242037139448441	0.00242037139448441\\
};\label{3x5_A10_sim}

\addplot [color=graphGreen, line width=1.3pt, mark=square,mark size = 3.0pt, mark options={solid, graphGreen}]
 table[row sep=crcr, y error plus index=2, y error minus index=3]{%
0.006	2.25e-05	9.99394832278014e-06	9.99394832278014e-06\\
0.01021429	0.0001	1.95990199754988e-05	1.95990199754988e-05\\
0.01942857	0.0007	0.000163927960275238	0.000163927960275238\\
0.02864286	0.0026	0.000315629383676489	0.000315629383676489\\
0.04707143	0.01074	0.000638871385020804	0.000638871385020804\\
0.0655	0.02762	0.00101574786301011	0.00101574786301011\\
0.08392857	0.04808	0.00132598474171372	0.00132598474171372\\
0.10235714	0.07708	0.00165313740657503	0.00165313740657503\\
0.13	0.12978	0.00208292909525639	0.00208292909525639\\
0.1574	0.1963	0.00246185916838474	0.00246185916838474\\
};\label{3x5_A100_sim}

\end{axis}

\node[draw,fill=white,inner sep=1.5pt,above left=0.5em] at (asymmetrical_3x5.south east){
  {\renewcommand{\arraystretch}{1.2}
    \begin{tabular}{lccc}
        \multicolumn{1}{c}{\textbf{}} & \textbf{$A =1$} & \textbf{$A =10$} & \textbf{$A =100$}\\
        bound & \ref{3x5_A1} & \ref{3x5_A10} & \ref{3x5_A100} \\
        simulation & \ref{3x5_A1_sim} & \ref{3x5_A10_sim} & \ref{3x5_A100_sim}\\
  \end{tabular}
}};

\end{tikzpicture}%
	}
	\caption{Logical error rate, $[[23,1,3/5]]$ surface code over symmetric and asymmetric channels. Comparison between the bound \eqref{eq:PeAsym} and the simulations with a \ac{MWPM} decoder.
		\label{Fig:plot_asymm_3x5}}
\end{figure}

\emph{4) [[23,1,3/5]] asymmetric surface code.} In Fig.~\ref{Fig:plot_asymm_3x5}, we compare the simulation \ac{MWPM} results for the $[[23,1,3/5]]$ surface code over symmetric and asymmetric channels with the error probability bound computed considering that no errors with weight $w > t$ could be corrected, i.e. all the $\beta_j$ would be zero for $j > t$. Specifically, we have determined these bounds using \eqref{eq:PeAsym} which accounts for the error rate of codes over channels with different error probabilities related to different kinds of errors ($\M{Z}$, $\M{X}$ and $\M{Y}$). It is interesting to note that, for all kind of bias of the channel $(A=1,10,100)$, the simulated logical error rate has the same behaviour of the bound error probability, but with a gap between each couple of curves. This gap is due to the capability of surface codes to correct many errors of weight $w \geq t + 1$. However, since not all the errors of weight $w = t + 1$ can be corrected, we have $\beta_{t+1}>0$ and this 
makes the asymptotic slope to be $t+1$, no matter how small is $\beta_{t+1}$. 


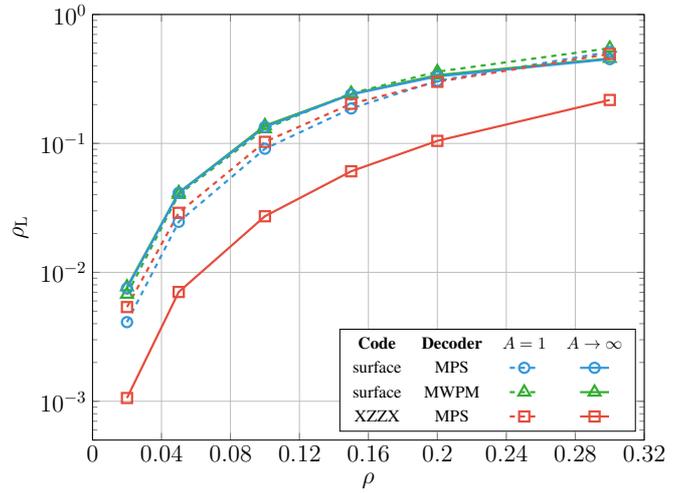
\begin{figure}[t]
	\centering
	\resizebox{0.49\textwidth}{!}{
%
%
\definecolor{mycolor1}{rgb}{0.00000,0.44700,0.74100}%
\definecolor{mycolor2}{rgb}{0.85000,0.32500,0.09800}%
\definecolor{mycolor3}{rgb}{0.92900,0.69400,0.12500}%
\definecolor{mycolor4}{rgb}{0.49400,0.18400,0.55600}%
\definecolor{mycolor5}{rgb}{0.46600,0.67400,0.18800}%
\definecolor{mycolor6}{rgb}{0.30100,0.74500,0.93300}%
\definecolor{mycolor7}{rgb}{0.92900,0.69400,0.12500}%
\begin{tikzpicture}

\begin{axis}[%
name = XZZX,
width=4.5in,
height=3.5in,
at={(0in,0in)},
scale only axis,
xmin=0,
xmax=0.32,
xminorticks=true,
xtick = {0, 0.04, 0.08, 0.12, 0.16, 0.2, 0.24, 0.28, 0.32},
xticklabels = {$0$, $0.04$, $0.08$, $0.12$, $0.16$, $0.2$, $0.24$, $0.28$, $0.32$},
xlabel={$\rho$},
xlabel style={font=\color{white!15!black}, font = \Large},
ymode=log,
ymin=0.0005,
ymax=1,
yminorticks=true,
ylabel={$\rho_\mathrm{L}$},
ylabel style={font=\color{white!15!black}, font = \Large},
axis background/.style={fill=white},
tick label style={black, semithick, font=\Large},
xmajorgrids,
ymajorgrids,
legend style={at={(0.03,0.97)}, anchor=north west, legend cell align=left, align=left, draw=white!15!black}
]

\addplot [color=graphGreen, dashed, line width=1.3pt, mark size = 4.0pt, mark=triangle, mark options={solid, graphGreen}]
 table[row sep=crcr, y error plus index=2, y error minus index=3]{%
0.02	0.00675	0.000507501399998069	0.000507501399998069\\
0.05	0.04014	0.0012166024937696	0.0012166024937696\\
0.1	0.12962	0.00208183608229121	0.00208183608229121\\
0.15	0.24463	0.00266434705728244	0.00266434705728244\\
0.2	0.36011	0.00297526959248509	0.00297526959248509\\
0.3	0.54748	0.0030850278564437	0.0030850278564437\\
}; \label{plot:XX_A1_MWMPFig6}

\addplot [color=brightBlue, dashed, line width=1.3pt, mark=o, mark size = 3.0pt, mark options={solid, brightBlue}]
 table[row sep=crcr, y error plus index=2, y error minus index=3]{%
0.02	0.00411	0.000507501399998069	0.000507501399998069\\
0.05	0.02462	0.0012166024937696	0.0012166024937696\\
0.1	0.09089	0.00208183608229121	0.00208183608229121\\
0.15	0.18657	0.00266434705728244	0.00266434705728244\\
0.2	0.30559	0.00297526959248509	0.00297526959248509\\
0.3	0.51265	0.0030850278564437	0.0030850278564437\\
}; \label{plot:XX_A1_MPSFig6}

\addplot [color=graphGreen, line width=1.3pt, mark size = 4.0pt, mark=triangle, mark options={solid, graphGreen}]
 table[row sep=crcr, y error plus index=2, y error minus index=3]{%
0.02	0.00771000000000001	0.000542128910365791	0.000542128910365791\\
0.05	0.04153	0.00123659161956792	0.00123659161956792\\
0.1	0.13727	0.00213295325776108	0.00213295325776108\\
0.15	0.24132	0.00265205207459461	0.00265205207459461\\
0.2	0.33852	0.00293296308125309	0.00293296308125309\\
0.3	0.45564	0.00308681140117215	0.00308681140117215\\
};  \label{plot:XX_Ainf_MWPMFig6}

\addplot [color=brightBlue, mark=o, mark size = 3.0pt, line width=1.3pt]
 table[row sep=crcr, y error plus index=2, y error minus index=3]{%
0.02	0.00751	0.000535105124025551	0.000535105124025551\\
0.05	0.04136	0.00123416751090215	0.00123416751090215\\
0.1	0.13411	0.00211211720685345	0.00211211720685345\\
0.15	0.23974	0.00264610694414991	0.00264610694414991\\
0.2	0.33069	0.00291595125194891	0.00291595125194891\\
0.3	0.45034	0.00308370908211368	0.00308370908211368\\
}; \label{plot:XX_Ainf_MPSFig6}

\addplot [color=brightRed, dashed, line width=1.3pt, mark=square, mark size = 3.0pt, mark options={solid, brightRed}]
 table[row sep=crcr, y error plus index=2, y error minus index=3]{%
0.02	0.00538	0.000453394036054291	0.000453394036054291\\
0.05	0.02895	0.00103920481829137	0.00103920481829137\\
0.1	0.10286	0.00188282276927129	0.00188282276927129\\
0.15	0.20295	0.00249283413101634	0.00249283413101634\\
0.2	0.30046	0.00284155246567337	0.00284155246567337\\
0.3	0.49414	0.00309881926060982	0.00309881926060982\\
}; \label{plot:XZ_A1Fig6}

\addplot [color=brightRed, mark=square, mark size = 3.0pt, line width=1.3pt]
 table[row sep=crcr, y error plus index=2, y error minus index=3]{%
0.02	0.00106	0.000201687371400393	0.000201687371400393\\
0.05	0.00704	0.000518212969322845	0.000518212969322845\\
0.1	0.02723	0.00100875331824664	0.00100875331824664\\
0.15	0.06079	0.00148099467513371	0.00148099467513371\\
0.2	0.10466	0.00189731932512964	0.00189731932512964\\
0.3	0.21723	0.00255583644599446	0.00255583644599446\\
}; \label{plot:XZ_AinfFig6}

\end{axis}

\node[draw,fill=white,inner sep=1.5pt,above left=0.5em] at (XZZX.south east){
  {\renewcommand{\arraystretch}{1.2}
    \begin{tabular}{cccc}
        \textbf{Code} &  
        \textbf{Decoder} &{$A = 1$} &  {$A \to \infty$ }\\
         {surface} & {MPS} & \ref{plot:XX_A1_MPSFig6} & \ref{plot:XX_Ainf_MPSFig6} 
         \\
          {surface} & {MWPM} & \ref{plot:XX_A1_MWMPFig6} & \ref{plot:XX_Ainf_MWPMFig6}   
         \\
         {XZZX} &  {MPS} & \ref{plot:XZ_A1Fig6} & \ref{plot:XZ_AinfFig6}  
         \\

  \end{tabular}
}};

\end{tikzpicture}%
	}
	\caption{ Logical error rate of surface codes and XZZX codes over depolarizing and  phase flip quantum channels. Comparison between a \ac{MWPM} and a \ac{MPS} decoders.
		\label{Fig:plot_mps}}
\end{figure}

\emph{ 5) Comparison between \ac{MWPM} and \ac{MPS} decoders. } In Fig.~\ref{Fig:plot_mps} we compare the performance of \ac{MWPM} decoders (for surface codes) and \ac{MPS} decoders (for both surface and XZZX codes). In particular, the latter is an approximation of a maximum likelihood decoder, hence we expect a better error correction capability \cite{BraSerSuc:14}. For surface codes, the gap between these two decoding techniques decreases as the channel bias increases. Indeed, the difference between their performance is particularly evident over a depolarizing channel, while they show almost the same results over a phase flip channel. Moreover, considering the \ac{MPS} decoder, we can osberve that the performance of the $[[13,1,3]]$ surface and the $[[13,1,3]]$ XZZX codes are quite similar over a depolarizing channel. However, over a channel with $\M{Z}$ errors only, the logical error rate of the surface code is higher since it is symmetrical and, as shown in Fig.~\ref{Fig:plot_asymmetrical}, it works better over symmetric channels. Contrarily, the performance of the XZZX codes are the best in this kind of scenario. Indeed, the intrinsic symmetries of this code make one kind of error ($\M{X}$ or $\M{Z}$) to align always on the same direction, orthogonal to the other. The result is that the logical error rate improves as the value of the bias increases. The figure provides a numerical evidence of the advantage of the XZZX codes, used with the \ac{MPS} decoder, over asymmetric channels. 

\section{Conclusions}\label{sec:conclusions}
We have assessed the performance of surface codes, providing analytical expressions and comparison with simulations, over symmetric and asymmetric quantum channels. For a surface code with minimum distance $d$,  the possibility to use simple decoders like the \ac{MWPM}, that allows to correct many patterns with more than $\lfloor (d-1)/2 \rfloor$ qubit errors, gives a considerable advantage with respect to bounded distance decoders. 
Our analysis emphasizes that, while for surface codes the performance degrades over a polarizing channel, the performance improves for XZZX codes of the same dimensions. 
Specifically, the behavior of logical error rate vs.\ physical error rate has been provided for the $[[13,1,3]]$, the $[[23,1,3/5]]$, the $[[33,1,3/7]]$, and the  $[[41,1,5]]$ surface codes. 

\section*{Acknowledgment}

This work was partially supported by the European Union under the Italian National Recovery and Resilience Plan (NRRP) of NextGenerationEU, HPC National Centre for HPC, Big Data and Quantum Computing (CN00000013).

\bibliographystyle{IEEEtran}
\bibliography{Files/IEEEabrv,Files/StringDefinitions,Files/StringDefinitions2,Files/refs}

\end{document}